\documentclass[prl,twocolumn,preprintnumbers,superscriptaddress,longbibliography, 10pt]{revtex4-2} 

\usepackage{graphicx}
\usepackage{bm}
\usepackage{amsmath}
\usepackage{amssymb}
\usepackage{amsfonts}
\usepackage{slashed}
\usepackage{latexsym,epsfig,bbm}
\usepackage{algorithm}
\usepackage{algorithmic}
\usepackage{bibunits}
\defaultbibliographystyle{apsrev4-2}
\defaultbibliography{ref}

\newcommand{\bra}[1]{\langle{#1}|}
\newcommand{\ket}[1]{|{#1}\rangle}

\newcommand{\braket}[2]{\langle{#1}|{#2}\rangle}

\newcommand{\figref}[1]{Fig.~\ref{#1}}
\newcommand{\figsref}[1]{Figs.~\ref{#1}}

\usepackage{color}
\definecolor{blue}{rgb}{0,0.2,1}

\definecolor{red}{rgb}{0.9,0,0}

\makeatletter
\def\maketitle{
\@author@finish
\title@column\titleblock@produce
\suppressfloats[t]}

\begin{document}
\bibliographyunit[\chapter]

\title{Spatiotemporal quantum advantages for rare-event sampling}
\let\thetitle\@title

\author{Akshat Gupta}
\email{akshat.gupta-4@postgrad.manchester.ac.uk}
\affiliation{Department of Physics \& Astronomy, University of Manchester, Manchester M13 9PL, United Kingdom}
\affiliation{Centre for Quantum Science and Engineering, University of Manchester, Manchester M13 9PL, United Kingdom}

\author{Jianjun Chen}
\email{chen1554@e.ntu.edu.sg}
\affiliation{Nanyang Quantum Hub, School of Physical and Mathematical Sciences,
Nanyang Technological University, Singapore 637371}
\affiliation{Centre for Quantum Technologies, Nanyang Technological University, Singapore 639798}

\author{Chengran Yang}
\email{yangchengran92@gmail.com}
\affiliation{Nanyang Quantum Hub, School of Physical and Mathematical Sciences,
Nanyang Technological University, Singapore 637371}
\affiliation{Centre for Quantum Technologies, Nanyang Technological University, Singapore 639798}

\author{Mile Gu}
\email{mgu@quantumcomplexity.org}
\affiliation{Nanyang Quantum Hub, School of Physical and Mathematical Sciences,
Nanyang Technological University, Singapore 637371}
\affiliation{Centre for Quantum Technologies, Nanyang Technological University, Singapore 639798}

\author{Caterina Doglioni}
\affiliation{Department of Physics \& Astronomy, University of Manchester, Manchester M13 9PL, United Kingdom}

\author{Jayne Thompson}
\affiliation{College of Computing and Data Science,
Nanyang Technological University, Singapore 639798}
\affiliation{Centre for Quantum Technologies, Nanyang Technological University, Singapore 639798}

\author{Thomas J.~Elliott}
\email{physics@tjelliott.net}
\affiliation{Department of Physics \& Astronomy, University of Manchester, Manchester M13 9PL, United Kingdom}
\affiliation{Department of Mathematics, University of Manchester, Manchester M13 9PL, United Kingdom}
\affiliation{Centre for Quantum Science and Engineering, University of Manchester, Manchester M13 9PL, United Kingdom}

\date{\today}
\makeatother
\begin{abstract}
In a world fraught with uncertainty, it behooves us to consider the most disastrous possible outcomes, in order that we can best mitigate against them. Often these are extreme risk events -- of high consequence, but low probability. Yet, the rarity of such events limits our ability to study them; there is a paucity of past occurrences to draw upon, and sampling such rare events from models is typically costly in time. Here, we introduce an approach for rare-event sampling based on quantum stochastic models. This integrates the memory advantages of quantum stochastic modelling -- and its associated efficient means of preparing quantum sample states -- with the `speed-ups' of quantum amplitude amplification, to yield simultaneous reductions in spatial and temporal computational complexity compared to classical Monte Carlo-based approaches.
\end{abstract}
\maketitle 

\begin{bibunit}

Global financial crises, major earthquakes, pandemics, and terror attacks are extreme risk events -- very rare, yet potentially catastrophic~\cite{hong2009granger, aven2013meaning, mcphillips2018defining, raymond2020understanding}. Their scarcity begets uncertainty, and unfamiliar warning signs often escape notice. Yet, early signals often exist: for example, in 2005, a small number of traders spotted surging subprime mortgage delinquencies -- ultimately leading to the 2008 crisis -- three years in advance~\cite{lewis2011big}. Identifying and monitoring such heralds of catastrophe is of paramount importance, to ensure we can handle these extreme occurrences. This is the remit of rare-event sampling~\cite{bucklew2004introduction, allen2009forward,bouchet2019rare}.

Given a model of a process, a common approach to rare-event sampling is to use Markov Chain Monte Carlo (MCMC)~\cite{geyer1992practical, brooks1998markov, chan2011rare, cerou2012sequential}, where many trajectories of the process are generated, and post-selected to keep only those trajectories where the rare event occurs for further study. However, if a target set of trajectories occurs with probability $p$, in order to reach a target number of samples from this set we must generate $O(1/p)$ trajectories. While more sophisticated approaches exist, such as a metadynamics biasing that reweights the model to favour rare event trajectories, these generically also involve a $O(1/p)$ scaling~\cite{chib1995understanding, robert1999monte, laio2008metadynamics, valsson2016enhancing}. This leads to a prohibitively harsh scaling in computational cost for sampling truly rare events.

\begin{figure}
	\includegraphics[width=\columnwidth]{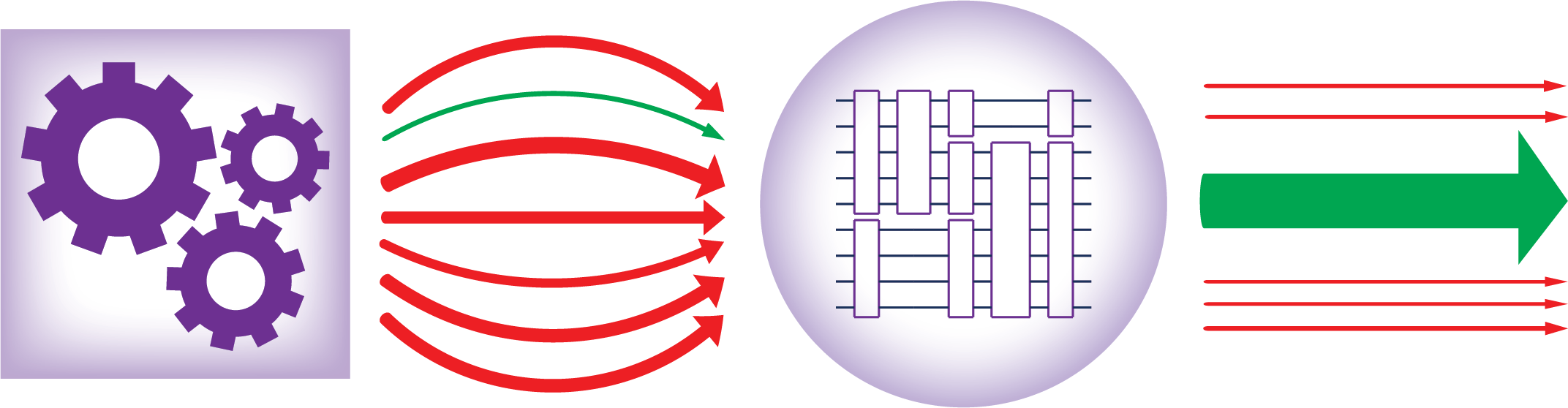}
	\caption{Conceptual schema of our approach. A quantum model is used to generate a weighted superposition over all trajectories of a stochastic process, including target rare event(s). This is then fed into an algorithm that amplifies the weighting for the target rare event(s) and suppresses the rest.}
	\label{fig:concept}
\end{figure}

With quantum computers on the horizon, promising computational speed-ups for a range of tasks, it is natural to ask whether a quantum algorithm can yield a better scaling. One may consider the use of the quantum amplitude amplification (QAA) primitive~\cite{brassard2000quantum} to boost the probability of rare events, which would seemingly allow for a quadratic reduction in temporal complexity. However, the direct naive and unstructured approach to constructing the quantum sample states that could be used as input into QAA would require an exponentially-deep circuit with increasing number of timesteps, and thus the complexity of state preparation would render any speed-up from QAA useless.

We overcome this limitation by incorporating the framework of quantum stochastic modelling~\cite{gu2012quantum, binder2018practical, liu2019optimal, loomis2019strong, elliott2021memory} with QAA. Specifically, we leverage that quantum models can generate the requisite quantum sample states for QAA with a circuit depth that grows only linearly with the number of timesteps, removing this barrier to the application of QAA. Moreover, quantum models of stochastic processes exhibit spatial complexity advantage, in terms of a reduced memory dimension. Many examples have demonstrated this quantum advantage can grow without bound; finite-sized quantum models can generate statistics inaccessible with any finite classical model~\cite{elliott2020extreme, wu2023implementing, Fanizza24_Quantum}.

Correspondingly, we show that rare-event trajectories can be sampled with $O(1/\sqrt{p})$ time complexity -- a quadratic reduction over the classical scaling -- while inheriting the spatial complexity advantages of the quantum models. The conceptual schema is illustrated in \figref{fig:concept}. Our approach is highly-related -- though methodologically distinct -- from Ref.~\cite{aghamohammadi2018extreme}, which explored the application of quantum stochastic models to rare-event sampling by examining quantum implementations of reweighted classical models, and showing that such `tilted' (i.e., biased) quantum models can exhibit (entropic) quantum memory advantages. By considering memory dimension our work introduces a stronger form of spatial complexity advantage, and through incorporating QAA, also yields a temporal complexity advantage.

We begin by introducing the relevant framework, on stochastic processes, quantum models, and QAA. We show how the quantum models can be deployed with QAA to achieve spatiotemporal advantages in rare-event sampling. We then demonstrate our approach with toy model applications. Finally, we conclude with a discussion of extensions, implications, and future work.

We consider rare-event sampling in the context of a time series manifest by an underlying stochastic process. Specifically, we consider discrete-time, discrete-event stochastic processes~\cite{khintchine1934korrelationstheorie}, where events $x\in\mathcal{X}$, drawn from a finite alphabet $\mathcal{X}$, occur at timesteps $t\in\mathbb{Z}$. We denote the random variable governing the event at timestep $t$ as $X_t$, with corresponding variate $x_t$. The full process $\mathcal{P}$ is defined by a joint distribution over all times, $P(\ldots, X_{0},X_{1},X_{2},\ldots)$. For shorthand, we will denote a contiguous sequence of events between timesteps $t_1$ and $t_2$ as $x_{t_1:t_2}=x_{t_1}x_{t_1+1}\ldots x_{t_2-1}$. For ease of notation we will assume stationary (i.e., time-invariant) processes, such that $P(x_{t:t+L})=P(x_{t':t'+L})\forall t,t',L\in\mathbb{Z}$, though our approach readily generalises by use of timestep-dependent evolution operators for the quantum models.

Such classical stochastic processes can be generated by hidden Markov models (HMMs). A HMM~\cite{rabiner1986introduction, upper1997theory} $\mathcal{M}$ is a triple $\mathcal{M}=\{\sigma_0,\mathcal{X},\{T^x\}_{x\in\mathcal{X}}\}$, where $\mathcal{X}$ is the event alphabet, (column) vector $\sigma_0$ the initial state, and $\{T^x\}$ a set of transition operators associated with the events~\footnote{An implicit state space for the HMM is defined according to an orthonormal basis that spans the reachable vectors $T^{w}\sigma_0$, for $w\in\mathcal{X}^*$.}. Mechanistically, the system undergoes a series of probabilistic transitions from the initial state according to the transition operators $\{T^x\}$, accompanied by the emission of an output event symbol $x\in\mathcal{X}$. The probability of a particular sequence $w=x_{0:L}$ occurring is $P(w)={\bm 1}T^w\sigma_0$, where ${\bm 1}$ is a (row) vector of ones, and we define $T^w=T^{x_{L-1}}\ldots T^{x_0}$.

Generalising to the quantum regime, a quantum HMM (QHMM)~\cite{monras2011hidden, clark2015hidden, monras2016quantum, adhikary2020expressiveness, elliott2021memory, zonnios2025quantum, riechers2025identifiability} is given by the triple $\mathcal{M}_q=\{\rho_0,\mathcal{X},\{\mathcal{A}_x\}_{x\in\mathcal{X}}\}$, where density operator $\rho_0$ is the initial state and $\{\mathcal{A}_x(\cdot)\}$ a set of completely-positive maps associated with each event~\footnote{As with HMMs, an implicit state space for a QHMM is defined according to the reachable states $\mathcal{A}_w(\rho_0)$, for $w\in\mathcal{X}^*$.}. Again, mechanistically the system undergoes transitions starting from its initial state, with the maps $\{\mathcal{A}_x\}$ playing the role of transition matrices; again, each transition is accompanied by emission of an output event $x\in\mathcal{X}$. To preserve probability normalisation, the overall channel $\sum_{x\in\mathcal{X}}\mathcal{A}_x(.)$ must be trace preserving. The probability of a particular sequence $w=x_{0:L}$ occurring is then given by $P(w)=\mathrm{Tr}(\mathcal{A}_w(\rho_0))$, where we have defined $\mathcal{A}_w(\cdot)=\mathcal{A}_{x_{L-1}}(\ldots (\mathcal{A}_{x_0}(\cdot))\ldots)$. As a strict generalisation, any HMM can be mapped to a corresponding QHMM~\cite{elliott2021memory}. Moreover, QHMMs are more expressive than HMMs of the same state space dimension~\cite{monras2016quantum, Fanizza24_Quantum}, leading to (sometimes unbounded) quantum advantages in memory cost for stochastic process generation, in both lossless and lossy simulation~\cite{ghafari2019dimensional, elliott2020extreme, elliott2021quantum, wu2023implementing, yang2025dimension}. These advantages underpin our claim of consequent spatial advantage in rare-event sampling with quantum stochastic models.

The approach we will introduce for rare-event sampling can be applied with any QHMM -- irrespective of whether it is defined through a mapping from a HMM, or learnt directly from data -- as QHMMs in general can be constructed as recurrent unitary circuits that assemble (generalised) quantum sample states. To avoid notational clutter we shall focus much of our exposition in terms of QHMMs as derived from a particular class of HMMs known as deterministic HMMs~\footnote{Note that \emph{deterministic} here refers to the update rule for the model's state given the current state and output; the model's output is still stochastic.}. We will cover the extension to general QHMMs later in the text.

Deterministic (or `unifilar') HMMs~\cite{crutchfield1989inferring, shalizi2001computational} are HMMs for which the current state can always be uniquely determined given the previous memory state and event. Given a deterministic HMM $\mathcal{M}$ with states $\{S_j\}$, a corresponding QHMM $\mathcal{M}_q$ always can be constructed with states $\{\ket{\sigma_j}\}$ and evolution operator $U$ according to the following implicit relation~\cite{binder2018practical, liu2019optimal}:
\begin{equation}
\label{eq:qmodel}
U\ket{\sigma_j}\ket{0}=\sum_x\sqrt{P(x|j)}e^{i\varphi_{xj}}\ket{\sigma_{\lambda(x,j)}}\ket{x},
\end{equation}
where $\lambda(x,j)$ is the update rule for the state, $P(x|j)=T^x_{\lambda(x,j)j}$, and $\varphi_{xj}\in[0,2\pi)$ are a set of tunable phases that we shall set here to zero for ease of notation. Measurement of the second subsystem yields the event for the timestep, and by iterating repeatedly with fresh ancillae for the second subsystem each time, events for arbitrarily many consecutive timesteps can be generated (see Supplementary Material).

If the event ancillae are not measured, then this procedure generates a superposition over all possible event trajectories of the corresponding length~\cite{yang2025dimension}:
\begin{equation}
\label{eq:qsample}
\mathcal{U}_L\ket{\sigma_j}\ket{0}^{\otimes L}=\sum_{x_{0:L}}\sqrt{P(x_{0:L}|j)}\ket{\sigma_{\lambda(x_{0:L},j)}}\ket{x_{0:L}},
\end{equation}
where $\mathcal{U}_L$ is the evolution corresponding to $L$ applications of $U$ across $L$ ancillary subsystems. This final state, the superposition over trajectories, is often referred to as a \emph{quantum sample state}~\cite{schuld2018quantum}, or `qsample' for short. Leveraging the recurrent structure of quantum models, these states are comparatively efficient to generate. Let $d\leq|\{S_j\}|$ be the dimension of the QHMM memory system; as noted above, $d$ can often be made substantially smaller than the dimension of the HMM state space. Then, using that a generic $n$ qubit gate decomposition into one and two qubit gates typically results in $O(4^n)$ gates~\cite{nielsen2000quantum}, the recurrent structure corresponds to a total depth $O(Ld^2|\mathcal{X}|^2)$ (see Supplementary Material A). Compared to the naive unstructured approach that does not leverage the recurrent structure, generically with depth $O(|\mathcal{X}|^{2L})$, this is an exponentially shorter depth with respect to the number of timesteps studied, and is core to making the prospect of temporal complexity advantage for rare-event sampling using QAA viable. As can be seen from the form of the qsample [Eq.~\eqref{eq:qsample}], the total space complexity involved is $\log_2 d +L\log_2|\mathcal{X}|$, the first term coming from the memory system, and the latter the $L$ timesteps of output events. The spatial quantum advantage is with respect to the first term, while the second term is identical for both classical and quantum approaches.

QAA~\cite{brassard2000quantum} generalises Grover's quantum search algorithm~\cite{grover1996fast}, inheriting the provable quadratic speed-up over optimal classical approaches. In QAA, the state space is partitioned into `good' and `bad' subspaces, the former containing all target states (as defined according to the particular problem at hand), and the latter the remainder of the state space. An initial state $\ket{\psi}$ can then without loss of generality be written as $\ket{\psi}=\sin(\theta)\ket{\mathrm{good}}+\cos(\theta)\ket{\mathrm{bad}}$. Following repeated application of the operator $\mathcal{G}=-R_\psi R_{\tau}$ to $\ket{\psi}$, where $R_\psi=I-2\ket{\psi}\bra{\psi}$, $R_{\tau}=I-2\Pi_\tau$, $I$ is the identity operator, and $\Pi_\tau$ the projection operator onto the good subspace, the resulting state is 
\begin{equation}
\mathcal{G}^n\ket{\psi}=\sin((2n+1)\theta)\ket{\mathrm{good}}+\cos((2n+1)\theta)\ket{\mathrm{bad}}.
\end{equation}
Given an appropriate $n$, this rotates the initial state towards its component in the good subspace. In the case that one has access to the operator $V$ that generates $\ket{\psi}$ from some fiducial state (i.e., $V\ket{0}=\ket{\psi}$), and its inverse, then we can replace $R_\psi=VR_0V^\dagger$, where $R_0=I-\ket{0}\bra{0}$.

We combine QAA with quantum stochastic models by using the quantum model evolution operator and its inverse as the state preparation operator for QAA. Specifically, let $U_{\mathrm{init}}$ represent the unitary operator that prepares the initial memory state from the fiducial $\ket{0}$ state, e.g., $\ket{\sigma_j}=U_{\mathrm{init}}\ket{0}$~\footnote{In the case where we wish to initialise in a distribution over memory states, e.g., $\rho = \sum_j P(j)\ket{\sigma_j}\bra{\sigma_j}$, then we instead use $U_{\mathrm{init}}$ to prepare a purification of $\rho$, such as $\sum_{j}\sqrt{P(j)}\ket{j}\ket{\sigma_j}$, and use the second subsystem for the memory in the remainder of the algorithm.}. Then, we can perform quantum model QAA by setting $V=\mathcal{U}_LU_{\mathrm{init}}$, with $L$ the desired length of the sequence we wish to sample over.

The target projection operator $\Pi_\tau$ is set according to the desired rare event, which we assume to have been determined in advance. For example, if this rare event is a particular sequence $x_{0:L}$, then $\Pi_\tau=I_{\mathrm{mem}}\otimes\ket{x_{0:L}}\bra{x_{0:L}}$, where $I_{\mathrm{mem}}$ is the identity operator over the memory subspace; $R_\tau$ can then be implemented as a single multi-controlled gate for each rare event; this is called once per Grover iteration. Likewise for $R_0$. These multi-controlled gates can be decomposed into at most $O((\log_2d+L\log_2|\mathcal{X}|)^2)$ elementary one and two qubit gates~\cite{barenco1995elementary}. The target rare event sequence can be shorter than $L$ by setting the remaining components of the projector on the output subspace to identity; similarly, the rare events can consist of particular non-contiguous outputs, interspersed by identity operators between target positions. Further, we can sample over multiple sequences in our rare event by taking a combined target projection operator that sums over individual projectors. 

If the total probability of the rare event(s) happening is $p$, then the amplitude of the `good' subspace is $\sqrt{p}$; note that we may neglect phase, as this can be rolled into $\ket{\mathrm{good}}$ without loss of generality. For $p\ll1$, this corresponds to $\theta\approx\sqrt{p}$. Thus, $n\sim O(1/\sqrt{p})$ Grover iterations are required to bring the target `good' subspace to approximately unit amplitude. Direct measurement can then be made to verify that we are in the target subspace. This corresponds to a quadratic reduction in query complexity compared to classical Monte Carlo-based methods, inherited from QAA. In the particular context of this problem and approach, this can also be reframed in terms of temporal complexity. In classical Monte Carlo, trajectories are generated and each checked to see if they contain rare events; the temporal complexity is linearly related to query complexity. Similarly, the gate depth (and temporal complexity) of QAA implemented in this manner is also linearly related to query complexity, with depth $O((Ld^2|\mathcal{X}|^2+(\log_2d+L\log_2|\mathcal{X}|)^2)/\sqrt{p}$)\footnote{The first term in the numerator corresponds to the depth in preparing (and inverting) the sample state, the second the cost of a multi-control gate, and the denominator coming from the requisite number of iterations.}. We can thus alternatively consider the advantage as a quadratic quantum reduction in temporal complexity. Coupled with the reduction in spatial complexity in the size of the memory inherited from the quantum models, this results in a quantum spatiotemporal advantage in rare-event sampling. We note again that use of the structure of quantum stochastic models results in a potentially exponential reduction in circuit depth -- for each call of the preparation unitary -- compared to a naive unstructured assembly of the quantum sample state for the future trajectories.

This approach also directly offers a quantum advantage in estimating the probability of rare events~\cite{suzuki2020amplitude, simon2024amplified}. Estimating the rare event probability corresponds to estimating the success probability of a binomial distribution, as each trajectory corresponds to either a rare event or not. Since the variance of a binomial distribution with success probability $p$ is $p(1-p)/n$ for $n$ trials~\cite{boas2006mathematical}, the confidence interval $\Delta p$ for an estimate of $p$ scales as $\sqrt{p(1-p)/n}$, with the prefactor dependent on the particular confidence interval. To constrain the estimate to a fixed proportional accuracy -- i.e., the relative error -- $\delta=\Delta p/p\propto\sqrt{(1-p)/np}$ thus requires $n\sim O(1/p\delta ^2)$ trials, for $p\ll1$. With QAA we can amplify the good subspace until its modified probability $p'\approx 1/2$. Then, the number of trials for fixed accuracy $\delta=\Delta p'/p'$ scales as $O(1/\delta^2)$. Considering that QAA requires $O(1/\sqrt{p})$ calls to the quantum sample state preparation unitary, this leads to a total query complexity $O(1/\sqrt{p}\delta^2)$, a quadratic advantage in $p$ compared to the classical approach. The spatial complexity reduction of the quantum models also applies.

\begin{figure}[ht!]
\raggedright
\includegraphics[width=0.6\columnwidth]{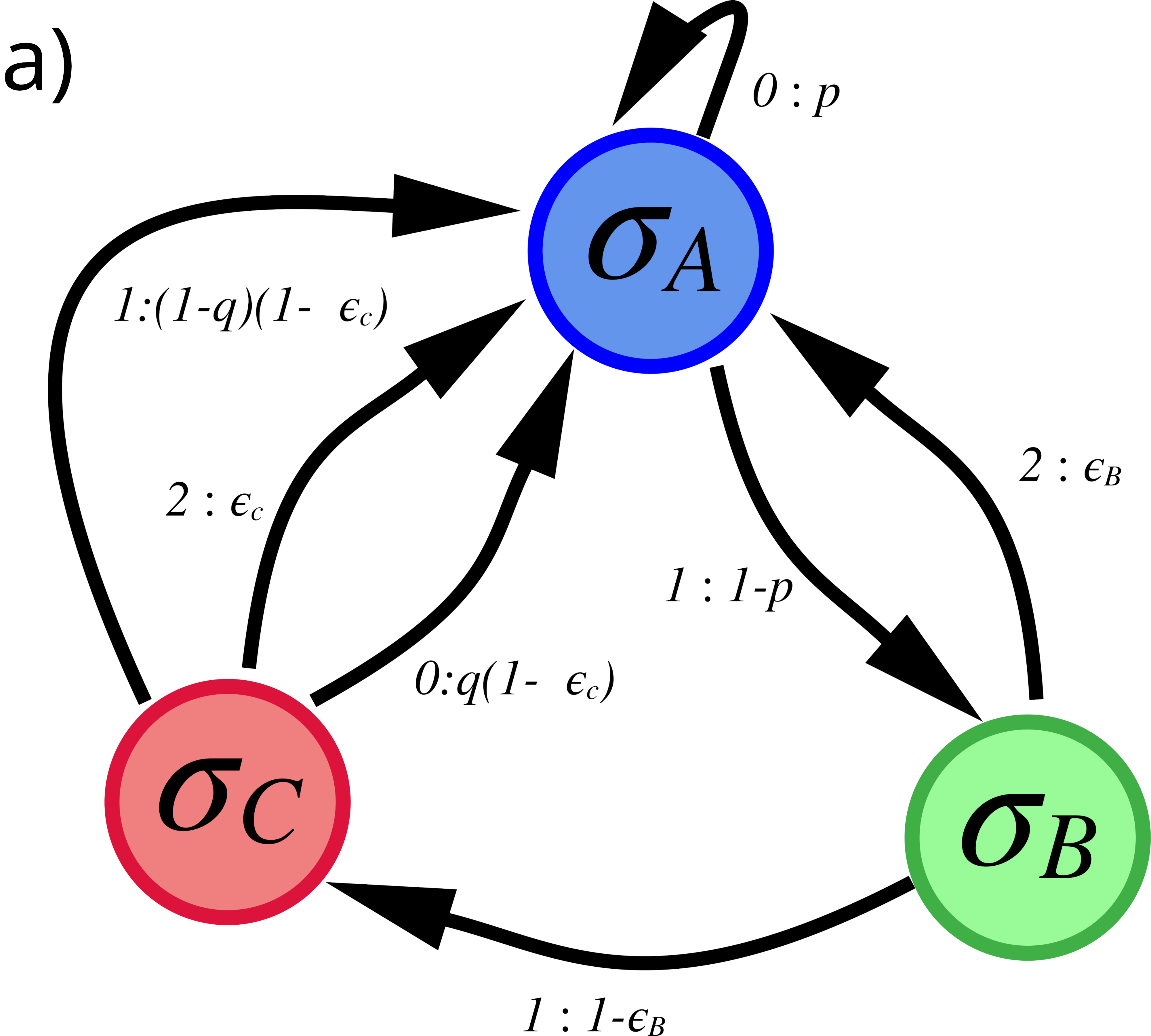}
\includegraphics[width=\columnwidth]{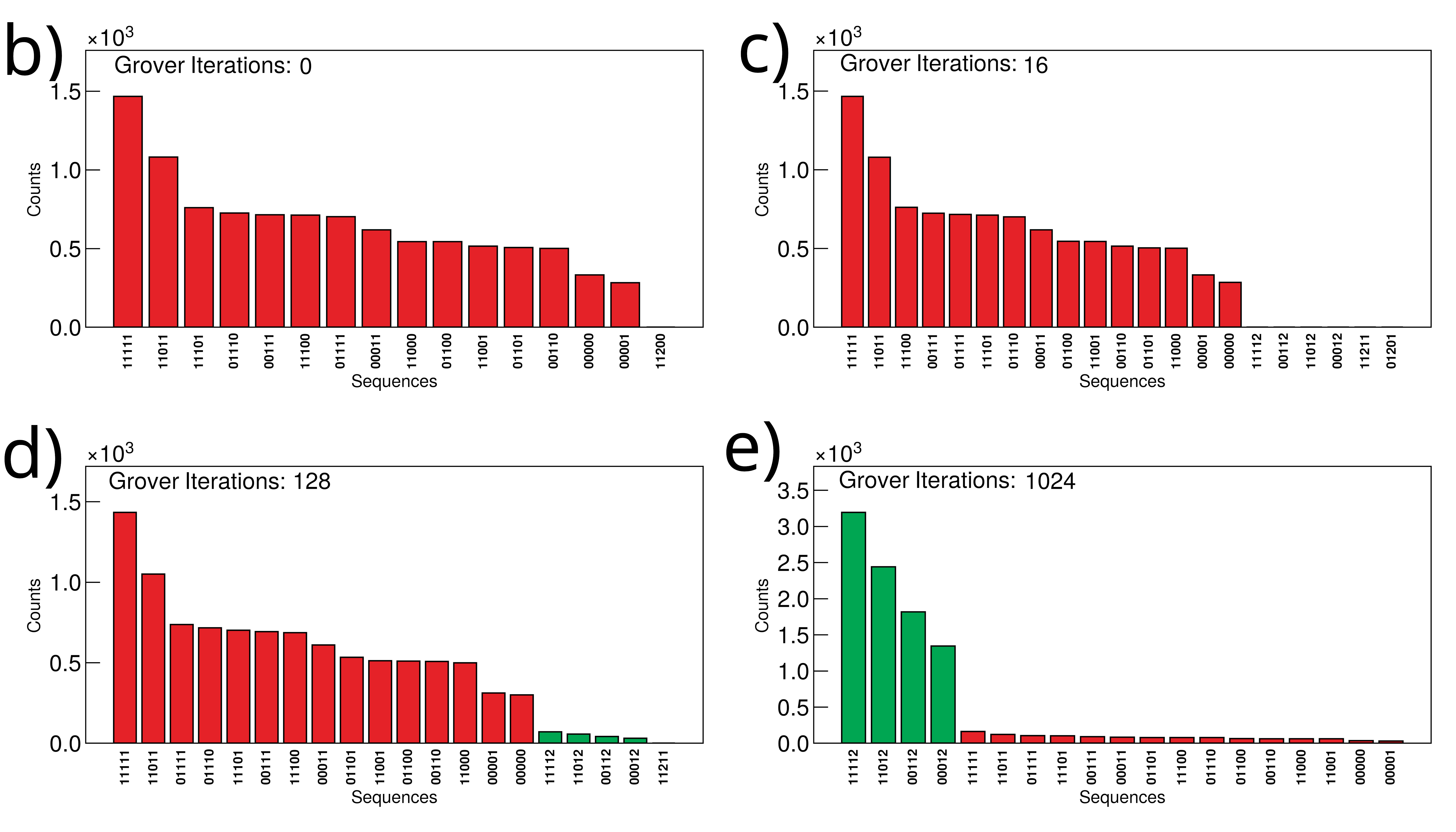}
\includegraphics[width=\columnwidth]{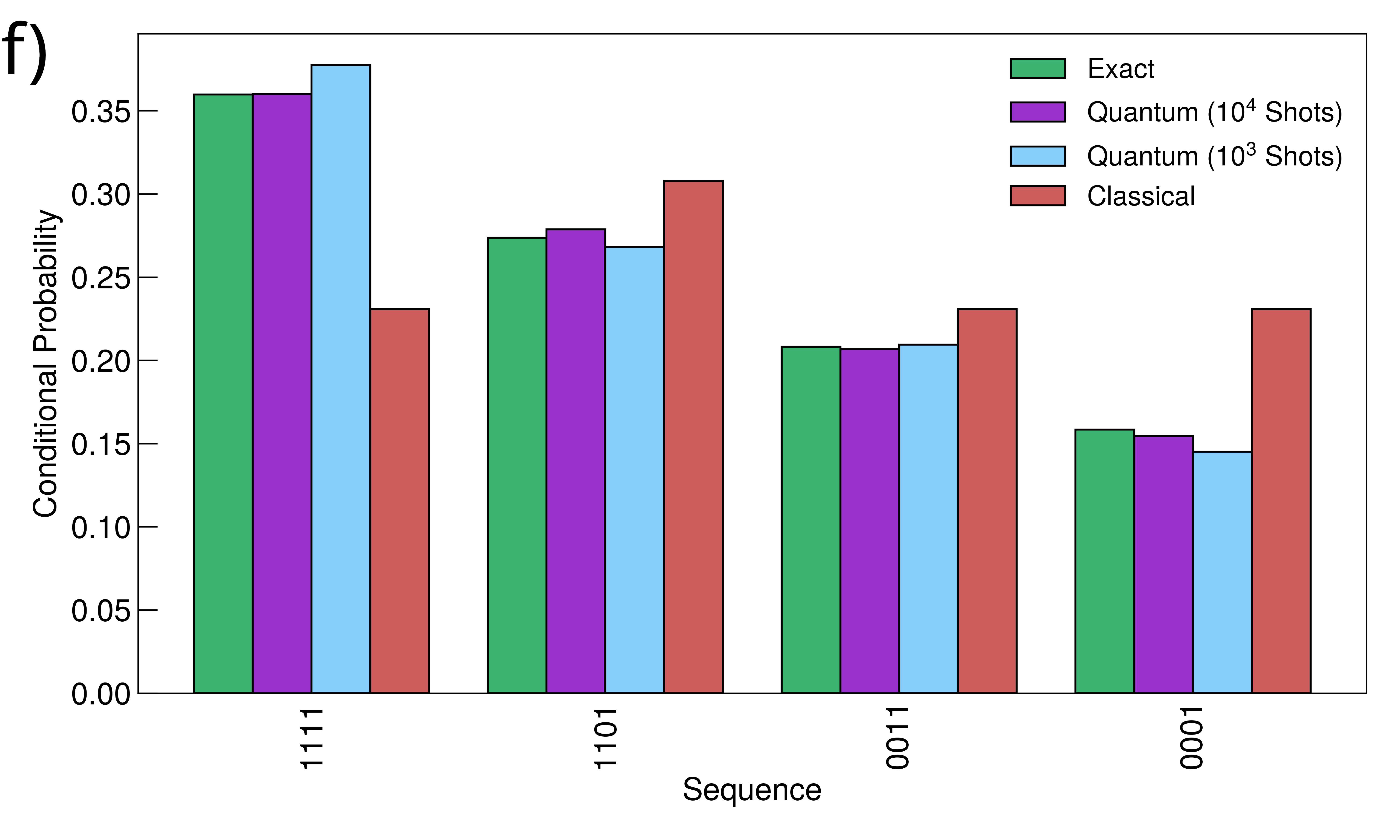}
	\caption{(a) Hidden Markov model representation of the Nemo process with error symbols; notation $x:p$ denotes the probability $p$ of the indicated transition occurring, accompanied by output symbol $x$. Progression of reweighting of rare events for (b) 0, (c) 16, (d) 128, and (e) 1024 Grover iterations; green denotes target rare events, red the remaining trajectories. (f) Comparison of reconstructed conditional probabilities for preceding symbols to a rare event for classical and quantum approaches; sequences with multiple errors omitted due to vanishingly-small probabilities of occurrence.}
	\label{fig:nemo}
\end{figure}

As an example, we apply our approach to a well-studied toy model, the Nemo process~\cite{mahoney2016occam}, modified to include an `error' symbol (`2') with small probability. The associated HMM is shown in \figref{fig:nemo}(a). For certain values of $p$ and $q$, this process admits a quantum model with reduced memory dimension $d=2$ (compared to the classical $d=3$). For particular ratios of $\epsilon_B$ and $\epsilon_C$, the error terms can be incorporated within the $d=2$ quantum model, with a free parameter controlling their absolute size. Details are given in the End Matter.

We take $p=0.5$, $q\approx 0.432$, and $\epsilon_B,\epsilon_C\sim O(10^{-6})$, and consider trajectories of length 5. Starting from state $\sigma_A$, we seek trajectories for which the final symbol is an error. Given these occur with probability $O(10^{-6})$, we require $O(10^3)$ Grover iterations to maximise the rare event trajectory probabilities. For $10^4$ shots, we show the progression of amplification in \figsref{fig:nemo}(b-e) with increasing numbers of Grover iterations, observing the expected reweighting in favour of the rare events.

We then use these amplified trajectories in order to sample the distribution of the preceding symbols when such a rare event occurs. We compare this with the exact values, calculated analytically [\figref{fig:nemo}(f)]. We also compare this with a classical Monte Carlo-based approach, which, for fair comparison, is given a number of shots equal to the number of times the quantum model evolution unitary is called, i.e., 2 $\times$ number of Grover iterations (1024) $\times$ number of shots of quantum circuit ($10^4$), corresponding to $2.048\times10^7$ shots in total. For the classical approach, only 13 rare event trajectories were generated from these shots; correspondingly, reconstruction of the distribution of symbols leading up to rare events is much poorer than that achieved by our quantum approach, for which 8796 trajectories were sampled from after amplification. Moreover, we see that even with an order of magnitude fewer shots of the quantum model a reasonably accurate reconstruction is achieved; in comparison, the analogous number of classical shots yielded only a single rare event trajectory.

In the End Matter, we also demonstrate application of our approach to a further toy model based on a renewal process -- representative of example application areas such as queueing theory~\cite{shortle2018fundamentals} and reliability analysis~\cite{arnold2020geometric} -- for which the corresponding QHMM has been experimentally implemented~\cite{wu2023implementing}. This latter example demonstrates a starker spatial complexity advantage ($d=2$, while $|\{S_j\}|=25$), a much rarer event ($p\sim O(10^{-15})$) corresponding to a greater temporal complexity advantage, and a larger number of timesteps.

The approach is readily extended to general HMMs, even when the property of determinism is not met. In such circumstances, to convert a HMM with transition matrices $\{T^x\}$ into a QHMM we must modify Eq.~\eqref{eq:qmodel} with an additional `junk' ancillary system. A valid approach for any HMM is as follows~\cite{elliott2021memory}:
\begin{equation}
U\ket{\sigma_j}\ket{0}\ket{0}=\sum_{xk}\sqrt{T^x_{kj}}\ket{\sigma_k}\ket{x}\ket{k},
\end{equation}
where the third subsystem is the junk, discarded at each timestep. Keeping the junk, along with the outputs, results in a modified quantum sample state after $L$ steps:
\begin{equation}
\label{eq:nondetL}
\mathcal{U}_L\ket{\sigma_j}\ket{0}^{\otimes 2L}\!=\!\!\!\sum_{x_{0:L}\bm{k}}\!\!\!\sqrt{T^{x_{L-1}}_{k_{L}k_{L-1}}\ldots T^{x_0}_{k_1j}}\ket{\sigma_{k_L}}\ket{x_{0:L}}\ket{\bm{k}},
\end{equation}
where $\bm{k}=(k_1,k_2,\ldots k_L)$ is the trajectory through the HMM's state space. The quantum model QAA approach then proceeds as above, but with the new quantum sample state preparation unitary as prescribed in Eq.~\eqref{eq:nondetL} above, and an extension of the projection operator $\Pi_\tau$ to include an identity operator for the junk subsystems carrying the state space trajectory. We remark that the specific assignment of the junk subsystems as prescribed above will generally negatively impact the quantum advantage in spatial complexity, while the the temporal/query complexity advantage remains unaffected. There often exists alternative junk encoding strategies that \emph{do} yield a spatial complexity advantage~\cite{elliott2021memory, sundar2026quantum}; Ref.~\cite{sundar2026quantum} provides stricter bounds on the dimension of the junk system based on the `maximal branching' of the underlying HMM. Moreover, we emphasise that nondeterminism is a property of a HMM, \emph{not} the underlying process, which permits multiple alternative, distinct HMM representations yielding identical statistics.

For a general QHMM $\mathcal{M}_q$, given the decomposition of the maps $\{\mathcal{A}_x\}$ in terms of their Kraus decomposition $\{\mathcal{A}_x(\cdot)=\sum_{y}K_{xy}\cdot K_{xy}^\dagger\}$, the corresponding unitary implementation is given by
\begin{equation}
U\ket{\cdot}\ket{0}\ket{0}=\sum_{xy}(K_{xy}\ket{\cdot})\ket{x}\ket{y}.
\end{equation}
Iteratively applying this with fresh ancillae at each timestep, one obtains the requisite $\mathcal{U}_L$ needed for input into QAA. Thus, our framework can be applied with \emph{any} QHMM.

In summary, we have introduced an approach that combines quantum stochastic modelling with QAA to realise a spatiotemporal reduction in complexity for rare-event sampling, incorporating memory advantages from the former with speed-ups from the latter. Such memory-efficient quantum models moreover enable a comparatively low-circuit depth means to assemble quantum sample states for QAA compared to generic unstructured approaches, making practical realisation of the approach more amenable. While our exposition is, for ease of presentation, in terms of HMMs with known transition matrices, the approach can be deployed with any QHMM. Correspondingly, lossy compression of quantum models~\cite{elliott2021quantum, yang2025dimension} may be incorporated -- thus further enhancing the spatial complexity advantage, as well as quantum models inferred directly from data. As remarked above, our approach is fully complementary to that of Ref.~\cite{aghamohammadi2018extreme}; using such biased quantum models could be employed as the input models for our approach, further integrating and enhancing quantum spatial and temporal complexity advantages.

QAA and QAE are central algorithms in proposals for the utility of quantum computing in enhancing computational power in a broad range of fields, such as finance~\cite{orus2019quantum, woerner2019quantum, egger2020quantum, wilkens2023quantum, ghosh2024energy}. By overcoming a significant obstacle in such algorithms -- that of efficiently preparing the input quantum sample state -- we offer a significant step-change towards the practical deployment of such approaches. One may readily expect that this approach readily generalises to other related quantum algorithms, such as quantum-enhanced rare event discovery and Monte Carlo integration~\cite{montanaro2015quantum, blank2021quantum, herbert2022quantum}. For very rare events, our approach may still require deep circuits. However, it may be possible to use circuit recompilation~\cite{jones2022robust} to overcome this, leveraging the repeating block nature of the circuits involved. We leave detailed exploration of this to future work.

An enticing further generalisation is to input-output processes~\cite{barnett2015computational}, where the observed behaviour is conditioned upon input stimuli presented to the system. Quantum models of input-output processes with memory advantages have been discovered~\cite{thompson2017using, elliott2022quantum}; using such a construction in place of the quantum stochastic models would allow quantum-enhanced rare event sampling to be performed for such processes. Notably, this offers the ability to input superpositions over inputs, which may be then sampled over with respect to a specific output string; this may find application in reinforcement learning~\cite{kaelbling1996reinforcement, sutton2018reinforcement}, by allowing an agent to deduce the input sequence(s) that have the highest probability to yield a target output.

In our motivation we have focussed on extreme risk events, such as natural disasters and global pandemics, i.e., catastrophic events that can lead to significant loss of life, substantial economic damage, or large-scale social upheaval. Yet, the importance of rare-event sampling stretches far beyond this, with applications across the sciences. For example, in high-energy physics, experiments at the Large Hadron Collider analyse proton-proton collisions where the probabilities of physical processes span over ten orders of magnitude, requiring simulation of enormous background processes to contextualise rare signals~\cite{10.3389/fdata.2023.1271639}. Monte Carlo generators~\cite{BUCKLEY2011145} employ sampling methods to simulate these interactions, generating collision events used to compare experimental data with theoretical predictions and search for evidence of new physics phenomena~\cite{PhysRevD.110.030001,CMS:2024zqs,ATLAS:2024lda}. Yet, these generators face a severe computational challenge: generating simulated datasets statistically comparable to data is prohibitive, especially if samples predominantly contain high-probability, known processes. This leads to a need for effective and efficient rare-event sampling algorithms that can bias samples towards generation toward low-probability, high-impact regions of phase space. Indeed, with many big data fields beset by bottlenecks due to limits of computational power, attention is turning to the prospect of quantum technologies to overcome such limits. Our approach may find further utility in such scenarios.

\acknowledgments
This work was funded by EPSRC Studentship EP/W524347/1 (Project 2932638) and the University of Manchester Dame Kathleen Ollerenshaw Fellowship.

\section*{End Matter}

Here we provide further details on the example in the Main Text. In the absence of the error symbols, the quantum model consists of three quantum memory states $\{\ket{\sigma_A},\ket{\sigma_B},\ket{\sigma_C}\}$, and an evolution operator $U$, defined implicitly through
\begin{align}
\label{eq:nemounitary}
U\ket{\sigma_A}\ket{0}&=\sqrt{p}\ket{\sigma_A}\ket{0}+\sqrt{1-p}\ket{\sigma_B}\ket{1}\nonumber\\
U\ket{\sigma_B}\ket{0}&=\ket{\sigma_C}\ket{1}\nonumber\\
U\ket{\sigma_C}\ket{0}&=\sqrt{q}e^{i\phi}\ket{\sigma_A}\ket{0}+\sqrt{1-q}e^{i\varphi}\ket{\sigma_A}\ket{1},
\end{align}
where $\phi$ and $\varphi$ are free, tunable parameters. Quantum dimension reduction is possible if the quantum memory states are linearly-dependent, i.e., we can express $\ket{\sigma_C}=\alpha\ket{\sigma_A}+\beta\ket{\sigma_B}$, for some $\alpha$ and $\beta$. By direct substitution into the right-hand side of Eq.~\eqref{eq:nemounitary}, it can be shown that this is satisfied when $\alpha=\sqrt{q/p}$ and $\beta=i(q(1-p)/p)^{1/4}$, under the condition that $(1-q)^2=q^3(1-p)/p^3$. For our particular choice of $p=0.5$, this requires that $q=1/12+\sqrt[3]{181/1728+\sqrt{13/864}}+\sqrt[3]{181/1728-\sqrt{13/864}}\approx 0.432$.

Without loss of generality, we can set $\ket{\sigma_A}=\ket{0}$, and $\ket{\sigma_B}=c_{AB}\ket{0}+\sqrt{1-|c_{AB}|^2}\ket{1}$, where $c_{AB}=\braket{\sigma_A}{\sigma_B}$, as can be calculated using Eq.~\eqref{eq:nemounitary}~\footnote{Specifically, we obtain $c_{AB}=(\alpha\beta^*+\sqrt{2}\beta)/(2-|\alpha|^2)$.}. Following the procedure of Refs.~\cite{binder2018practical, liu2019optimal}, we can determine the action of $U$ on the basis states. We then introduce the error symbol by modifying $U$ into a new unitary operator $U_{\epsilon}$ as follows:
\begin{align}
U_\epsilon\ket{0}\ket{0}&=U\ket{0}\ket{0}\nonumber\\
U_\epsilon\ket{1}\ket{0}&=\sqrt{1-\delta}U\ket{1}\ket{0}+\sqrt{\delta}\ket{0}\ket{2},
\end{align}
where the remaining columns can be set arbitrarily, provided they maintain orthogonality with each other and the prescribed columns. Here, $\delta$ is a parameter that controls the strength of the error; for our simulation, we set $\delta=10^{-6}$. We can relate this to the error probabilities as follows: $\epsilon_B=(1-|c_{AB}|^2)\delta$ and $\epsilon_C=|\beta|^2\epsilon_B$. For our parameters, this corresponds to $\epsilon_B\approx 0.88\times10^{-6}$ and $\epsilon_C\approx 0.58\times 10^{-6}$.

As a second example, we consider a periodically-modulated decay process~\cite{wu2023implementing}. This is a renewal process~\cite{marzen2015informational}, which can be represented as a deterministic HMM with alphabet $\mathcal{X}=\{0,1\}$, $N$ states $\{S_n\}$ $n=0,1,\ldots,N-1$, with transitions satisfying $P(0|S_n)=\Phi(n+1)/\Phi(n)$, where emissions of 0 are accompanied by a transition $S_n\to S_{(n+1)\mathrm{mod}N}$ and emissions of 1 $S_n\to S_0$. The `survival probability' $\Phi(n)=\Gamma^n(1-V\sin^2(n\pi/N))$.  This process permits a significant quantum advantage in spatial complexity -- while the minimal classical model requires a $\log_2 N$ bit memory, a quantum model can be constructed that requires only a single qubit of memory. We consider the case where $N=25$, $V=1/3.05$, and $\Gamma\approx 0.9511$. Specifically, we follow the procedure detailed in the Supplementary Material of Ref.~\cite{wu2023implementing}, resulting in Kraus operators
\begin{align}
K_0 &= \begin{pmatrix}0.97525 & -0.0061278 + 0.048507i\\ 0  0.94461 + 0.24253i\end{pmatrix} \nonumber\\
K_1 &=\begin{pmatrix} -0.16987 & -0.020763 + 0.16436i \\ -0.14156 &-0.017302 + 0.13696i \end{pmatrix}
\end{align}
(up to 5 significant figures).

\begin{figure}
\includegraphics[width=\columnwidth]{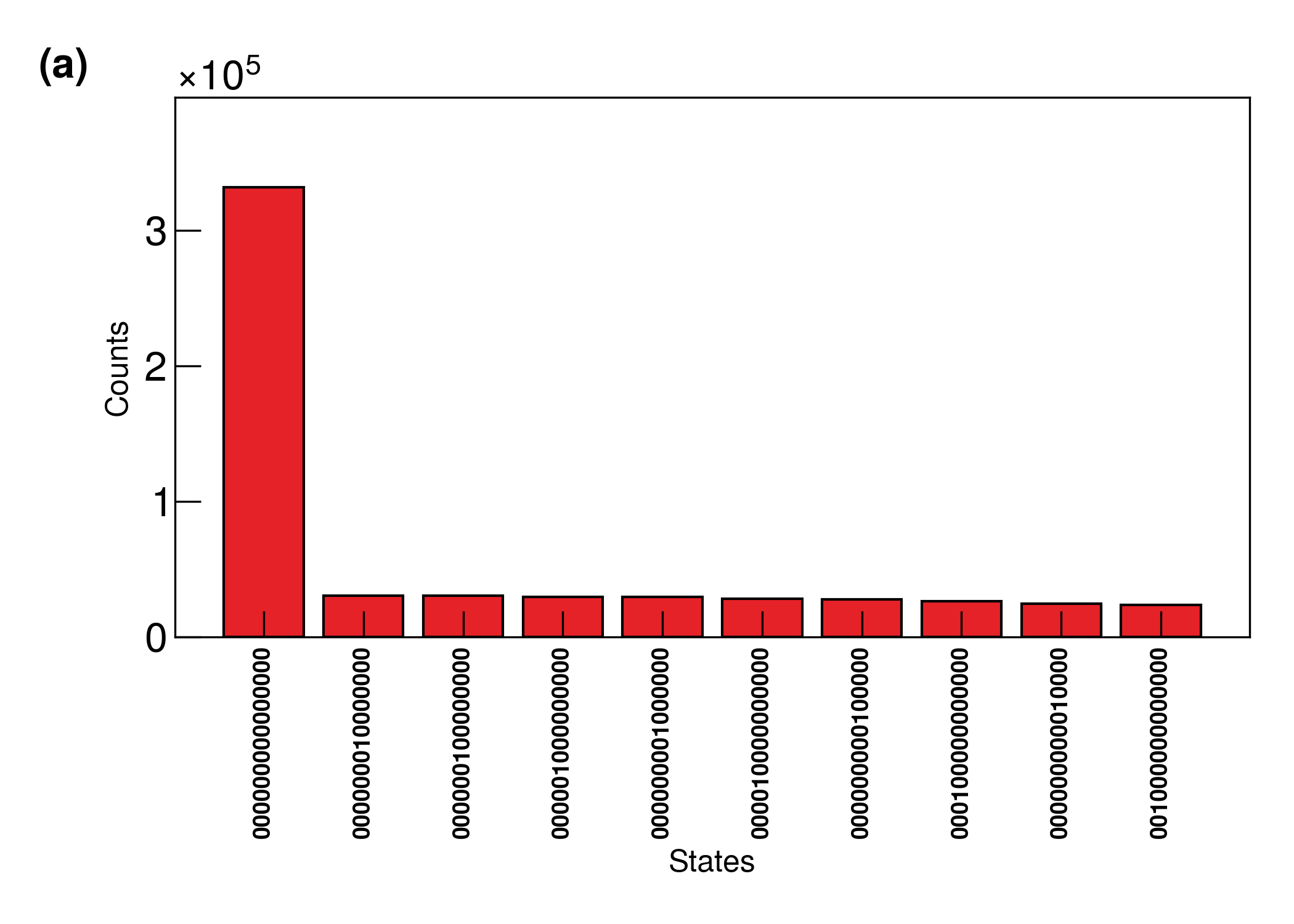}
\includegraphics[width=\columnwidth]{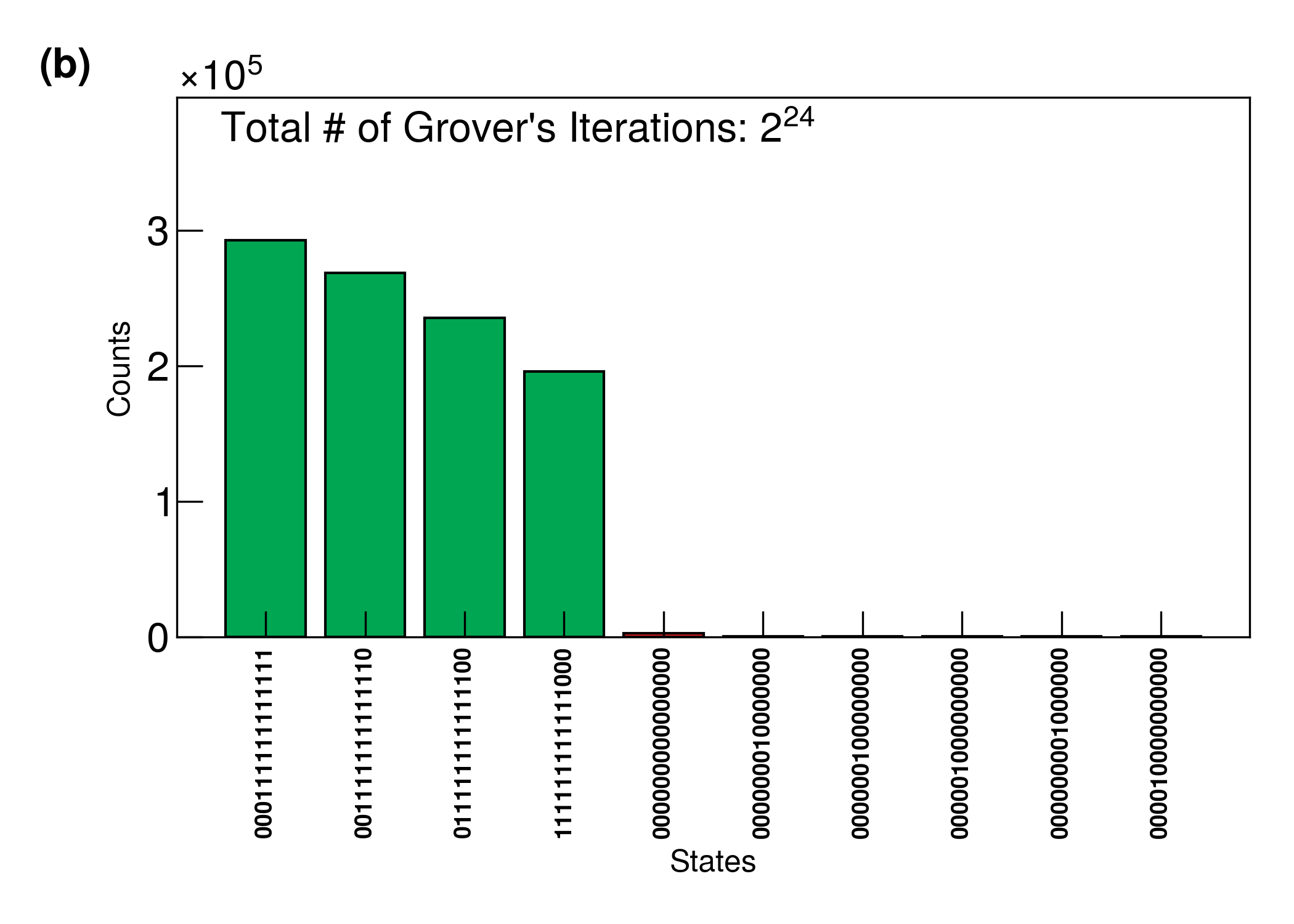}
	\caption{Sampled trajectories for the PMD process (a) without amplification, and (b) with $2^{24}$ Grover iterations; green denotes target rare events, red the remaining trajectories. Circuits evaluated with $10^6$ shots. Plots limited to 10 highest-probability sequences. Plots of the progression of amplification are given in the Supplementary Material.}
	\label{fig:pmdexample}
\end{figure}

We consider 15 timesteps of evolution, starting from the $S_0$ state. As our rare event, we consider sequences containing exactly 12 consecutive 1s, with the remaining events all 0. There are four such sequences (000111111111111, 001111111111110, 011111111111100, 111111111111000), with total probability $p\approx2.445\times10^{-15}$. Thus, classical Monte Carlo sampling has a below 25\% probability of yielding a single rare event sequence in $10^{14}$ trajectories. In contrast, we see that after $2^{24}\approx 1.68\times10^7$ rounds of amplitude amplification, sampling from the quantum state with high probability $(\approx0.992)$ yields a rare event sequence [\figref{fig:pmdexample}]. Thus, there is a substantial quantum advantage also in query and temporal complexity.

\putbib[ref,fullNotes]

\end{bibunit}


\clearpage

\pagebreak
\widetext

\title{Supplementary Material\\ \thetitle}
\maketitle
\onecolumngrid

\setcounter{equation}{0}
\setcounter{figure}{0}
\setcounter{table}{0}
\setcounter{page}{1}
\renewcommand{\theequation}{S\arabic{equation}}
\renewcommand{\thefigure}{S\arabic{figure}}
\renewcommand{\thepage}{S\arabic{page}}
\renewcommand{\bibnumfmt}[1]{[S#1]}
\renewcommand{\citenumfont}[1]{S#1}

\begin{bibunit}

\section{A: Linear scaling of circuit depth with number of timesteps}

Here we provide further details of the claim in the Main Text that the QHMM formalism allows for the quantum sample state to be prepared with a circuit depth that scales only linearly with the number of timesteps.

Let us first show, schematically, how QHMM evolution takes place over multiple timesteps:
\begin{center}
\includegraphics[width=0.9\columnwidth]{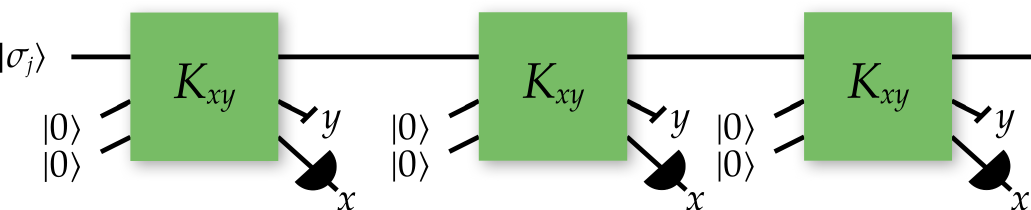}
\end{center}
The top branch contains the persistent memory that is carried over each timestep. At each timestep, this memory system is coupled with ancillae initialised in the fiducial state $\ket{0}$. For non-deterministic QHMMs it is convenient to split this into $\ket{0}\ket{0}$, in order to make apparent the division into component Hilbert spaces in the inputs and outputs. The coupling unitary $U$ acts on the memory and these ancillae, to produce a weighted superposition over possible outputs (conditioned on the memory state) and update the memory state accordingly. Measurement of the first output space gives the output $x$ for that timestep, and the second output, containing the auxilliary index $y$, may be discarded (however, it must be kept if the quantum sample state is to remain coherent). In terms of the Kraus operators of the QHMM $\{K_{xy}\}$, $U$ acts as
\begin{equation}
U\ket{\cdot}\ket{0}\ket{0}=\sum_{xy}(K_{xy}\ket{\cdot})\ket{x}\ket{y}.
\end{equation}

To simplify notation, we will follow the bulk of the Main Text and continue this exposition explicitly for deterministic QHMMs constructed from mappings of deterministic HMMs. However, we emphasise that this procedure is fully general for any QHMM; one can simply replace the output spaces containing $\{\ket{x}\}$ below with the composite output space containing $\{\ket{x}\ket{y}\}$. Correspondingly, $U$ takes the specific form prescribed in Eq.~\eqref{eq:qmodel} of the Main Text:
\begin{equation}
U\ket{\sigma_j}\ket{0}=\sum_x\sqrt{P(x|j)}e^{i\varphi_{xj}}\ket{\sigma_{\lambda(x,j)}}\ket{x},
\end{equation}
where $P(X|J)$ and the update rule $\lambda(x,j)$ are determined by the original HMM, $\{\varphi_{xj}\}$ are a set of phase parameters that may be set arbitrarily, but can be tuned to vary the quantum memory advantage. $U$ and the memory states $\{\ket{\sigma_j}\}$ are defined implicitly (up to a local isometry) by this equation, and may systematically be constructed using the procedure given in Refs.~\cite{binder2018practical, liu2019optimal}. We will set $\{\varphi_{xj}=0\}$ from hereon, to simplify notation.

Let us now consider the application of this primitive over multiple timesteps. Consider an initial memory state $\ket{\sigma_j}$, and $L$ ancillae initialised in $\ket{0}$, where $L$ is the target number of timesteps. Let $U_{[n]}$ denote the application of $U$ to couple the memory system with the $n$th ancilla. Thus, we can represent the action of $U_{[1]}$ as
\begin{equation}
U_{[1]}\ket{\sigma_j}\ket{0}^{\otimes L}=\sum_{x_1}\sqrt{P(x_1|j)}\ket{\sigma_{\lambda(x_1,j)}}\ket{x_1}\ket{0}^{\otimes L-1}.
\end{equation}
We can then perform a second timestep of evolution using $U_{[2]}$ to couple the memory to the second ancilla:
\begin{equation}
U_{[2]}U_{[1]}\ket{\sigma_j}\ket{0}^{\otimes L}=\sum_{x_1x_2}\sqrt{P(x_1x_2|j)}\ket{\sigma_{\lambda(x_1x_2,j)}}\ket{x_1}\ket{x_2}\ket{0}^{\otimes L-2},
\end{equation}
where $\lambda(x_1x_2,j)$ is the two timestep concatenation of the update rule.

Repeating this procedure for all $L$ timesteps, we have that 
\begin{equation}
U_{[L]}\ldots U_{[1]}\ket{\sigma_j}\ket{0}^{\otimes L}=\sum_{x_1\ldots x_L}\sqrt{P(x_1\ldots x_L|j)}\ket{\sigma_{\lambda(x_1\ldots x_L,j)}}\ket{x_1}\ldots\ket{x_L}.
\end{equation}
Denoting $\mathcal{U}_L:=U_{[L]}\ldots U_{[1]}$ and using $x_{0:L}$ to denote the concatenated string of $L$ outputs, we thus obtain Eq.~\eqref{eq:qsample} of the Main Text.

Graphically, a single timestep is represented by the circuit
\begin{center}
\includegraphics[width=0.7\columnwidth]{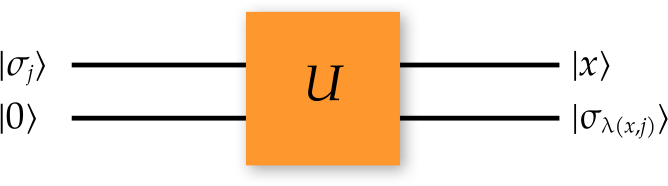}
\end{center}
Note that we have exchanged the order of the outgoing subspaces such that the memory and output subspaces change places. While this is not fundamentally necessary, it assists with the pedagogy of this diagrammatic exposition, and also may have practical consequences in terms of connectivity on a real quantum processor. This incurs, at worst, a simple swap circuit between the two subspaces. Nevertheless, from a theoretical algorithmic standpoint this is not needed.

For two timesteps, we perform two such blocks, using $U_{[1]}$ and $U_{[2]}$. The former leaves the second ancilla untouched; the the latter leaves the first output subspace untouched.
\begin{center}
\includegraphics[width=0.7\columnwidth]{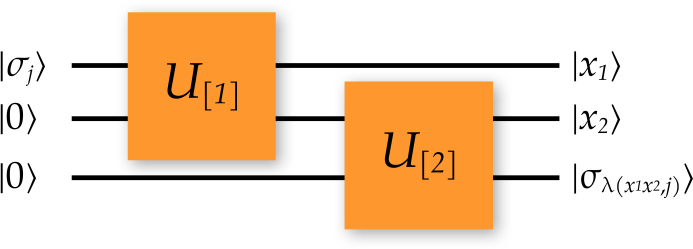}
\end{center}

Iterating, we can sequentially apply $U_{[1]}$ through to $U_{[L]}$ to generate $L$ timesteps:
\begin{center}
\includegraphics[width=0.9\columnwidth]{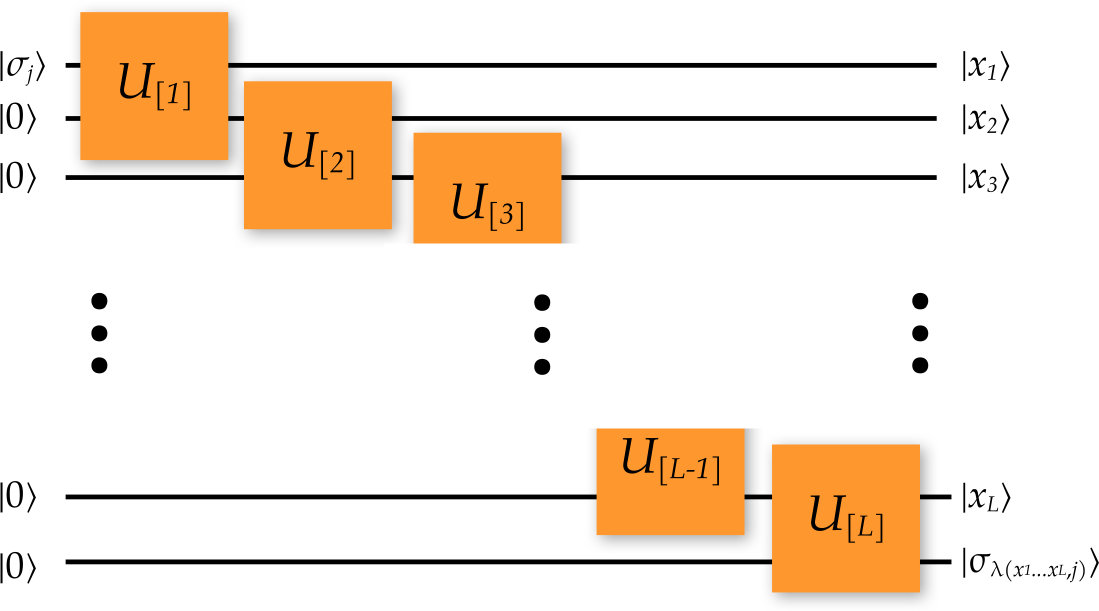}
\end{center}

Crucially, the $U_{[n]}$ are all an identical series of gates; they differ only in terms of the subspaces they act on in the circuit. Thus, given the ability to implement one such $U$ for a single timestep, the $L$-timestep concatenation $\mathcal{U}_L$ consists of simply applying this primitive $L$ times. Thus, if the gate complexity (or depth) of applying a single timestep is $C$, the complexity (or depth) of applying $L$ timesteps is $LC$ -- a linear scaling. 

As remarked in the Main Text, decomposing an $n$ qubit gate into one and two qubit gates typically results in $O(4^n)$ gates~\cite{nielsen2000quantum}. Thus, the depth of each $U$ is $O(d^2|\mathcal{X}|^2)$, corresponding to the square of the dimensions of the memory and single-timestep ancilla subspaces. Correspondingly, the total depth of $\mathcal{U}_L$ scales as $O(Ld^2|\mathcal{X}|^2)$. This should be compared to a direct, unstructured decomposition of a gate acting over the entire set of $L$ ancillae needed for assembling the full qsample, which would generically result in depth $O(|\mathcal{X}|^{2L})$, exponentially worse than our QHMM-based recurrent approach.

\section{B: Progression of amplification for PMD process}

\begin{figure}
\includegraphics[width=\columnwidth]{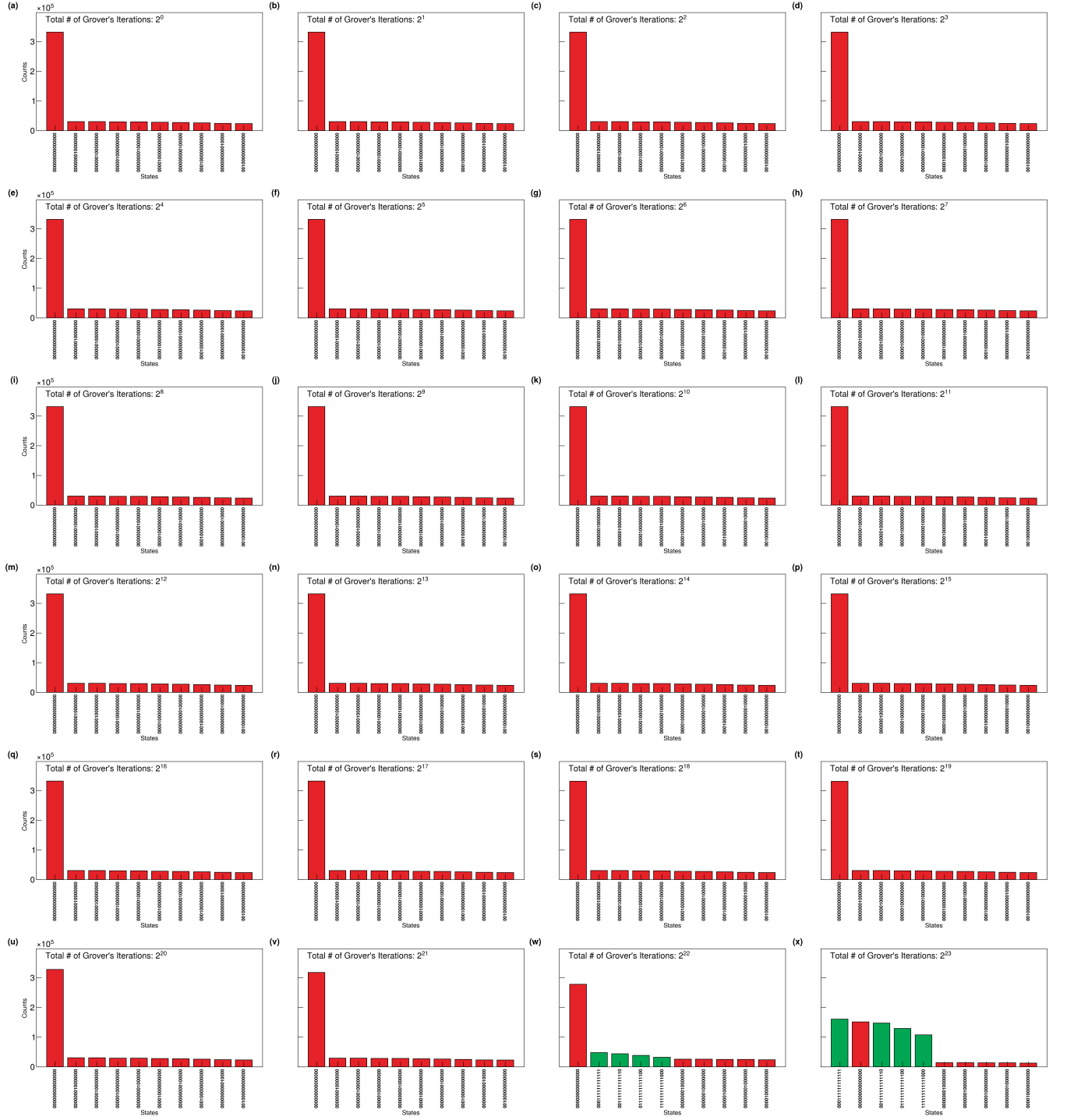}
	\caption{(a-x) Progression of amplification of target rare event sequences for the PMD process with increasing number of iterations; green denotes target rare events, red the remaining trajectories. Plots limited to 10 highest-probability sequences.}
	\label{fig:pmdprog}
\end{figure}

In \figref{fig:pmdprog} we display the progression of amplification of the target rare events with increasing number of cycles for the PMD process discussed in the End Matter.

\putbib[ref]
\enlargethispage{100em}
\vspace{50em}

\end{bibunit}

\end{document}